\documentstyle[vsolj01,graphicx]{article}

\begin{document}

\title{Reanalysis of Ichinohe's Observations of RU Cam}

\author{Taichi Kato}
\author{(Dept. of Astronomy, Kyoto University, Kyoto 606-8502, Japan)}

\author{Seiichi Sakuma}
\author{(Kawasaki, Japan)}

The W Vir-type variable star, RU Cam, once claimed to have stopped
pulsation (Demers, Fernie 1966; Huth 1966), has been known to show
irregularly variable amplitude (Kollath, Szeidl 1993).  Sato et al. (1997)
has also shown from the analysis of the VSOLJ archive that visual observers
of the VSOLJ have recorded the cessation of pulsation and its possible
resumption, together with the phase of large-amplitude variation around
1910.  The observations around 1910 were done by Ichinohe, a part of which
was published by the observer himself (Ichinohe 1909).  We present the
result of reanalysis of Ichinohe's original data until 1911, by using the
Hipparcos {\it V}-magnitudes of the comparison stars.

The original observational records using Pogson's step method were
first analyzed by a least-squares program (Kato, in preparation)
to obtain the best step values of the comparison stars and the variable
star.  The resultant step values of the comparison stars were used to
calibrate the original observation into the Hipparcos {\it V} system.
Table 1 summarizes the result of best determined step values of the
comparison stars.

A least-squares fit of the step values to the Hipparcos {\it V}-magnitudes
yielded the following conversion equation.
\begin{equation}
V = 8.70 + 0.191\cdot step
\end{equation}
By using this, we have reduced the original observations to the present
Hipparcos {\it V}-scale.  The result is tabulated in Table 2.

\begin{table}[b]
\begin{center}
Table 1. Comparison stars \\
\vspace{6pt}
\begin{tabular}[tbhp]{c c c c c}
\hline
Desig. &        BD        & Steps & Hipparcos V &  O-C$^*$ \\
\hline
a      & +70$^{\circ}$450 &  0.00 &    8.75     &  0.05 \\
c      & +70$^{\circ}$447 &  1.33 &    9.05     &  0.10 \\
d      & +69$^{\circ}$422 & -3.40 &    8.04     & -0.01 \\
e      & +69$^{\circ}$420 &  1.44 &    8.90     & -0.07 \\
b      & +70$^{\circ}$448 &  2.29 &    9.06     & -0.08 \\
g      & +70$^{\circ}$445 &  5.29 &    9.85     &  0.14 \\
f      & +70$^{\circ}$453 &  4.54 &    9.43     & -0.14 \\
\hline
\multicolumn{5}{l}{$^*$ Residuals from eq. (1)}
\end{tabular}
\end{center}
\end{table}

\begin{table}
\begin{center}
Table 2. Ichinohe's observations \\
\vspace{6pt}
\begin{tabular}[tbhp]{c c c c c c}
\hline
JD-2410000 & mag & JD-2410000 & mag & JD-2410000 & mag \\
\hline
7649.547 & 8.69 & 7743.796 & 9.13 & 8364.126 & 8.89 \\
7650.805 & 9.12 & 7746.817 & 8.90 & 8368.954 & 9.06 \\
7651.823 & 9.08 & 7747.822 & 8.74 & 8390.133 & 9.09 \\
7652.708 & 9.21 & 7749.846 & 8.25 & 8392.944 & 8.75 \\
7654.726 & 9.47 & 7759.819 & 8.37 & 8405.003 & 8.99 \\
7657.799 & 8.78 & 7762.650 & 9.30 & 8406.051 & 9.57 \\
7658.645 & 9.06 & 7763.797 & 9.18 & 8413.047 & 8.99 \\
7658.646 & 9.06 & 7881.059 & 8.64 & 8422.085 & 8.61 \\
7664.665 & 8.37 & 7889.149 & 8.43 & 8424.165 & 8.25 \\
7665.646 & 8.47 & 7890.073 & 8.24 & 8445.040 & 8.65 \\
7666.584 & 8.48 & 7904.153 & 8.64 & 8446.042 & 8.56 \\
7667.588 & 8.71 & 7945.240 & 9.07 & 8450.061 & 8.94 \\
7668.605 & 8.74 & 7949.206 & 8.48 & 8450.961 & 9.35 \\
7668.653 & 8.70 & 7950.060 & 8.44 & 8526.285 & 8.47 \\
7671.611 & 8.88 & 7950.988 & 8.44 & 8538.253 & 8.35 \\
7671.726 & 8.95 & 7957.108 & 8.32 & 8558.232 & 8.55 \\
7671.826 & 9.07 & 7962.941 & 8.95 & 8664.018 & 7.80 \\
7675.591 & 9.53 & 7966.037 & 9.18 & 8714.062 & 8.56 \\
7676.586 & 9.79 & 8057.105 & 9.12 & 8738.181 & 8.96 \\
7679.625 & 9.41 & 8117.068 & 8.86 & 8920.202 & 8.74 \\
7680.709 & 9.01 & 8148.997 & 8.44 & 8951.024 & 8.97 \\
7682.697 & 8.61 & 8168.163 & 9.07 & 8971.126 & 8.06 \\
7685.835 & 8.00 & 8173.192 & 8.22 & 8982.881 & 8.93 \\
7686.631 & 8.50 & 8174.209 & 8.32 & 9028.997 & 9.43 \\
7687.600 & 8.38 & 8175.125 & 8.42 & 9053.913 & 8.99 \\
7689.715 & 8.37 & 8181.199 & 8.47 & 9061.997 & 8.41 \\
7703.648 & 8.76 & 8183.074 & 8.54 & 9062.985 & 8.41 \\
7704.697 & 8.55 & 8211.088 & 9.05 & 9072.997 & 9.24 \\
7705.628 & 8.37 & 8212.120 & 8.91 & 9088.180 & 8.31 \\
7706.709 & 8.19 & 8215.051 & 8.26 & 9090.235 & 8.54 \\
7708.680 & 8.22 & 8216.942 & 8.43 & 9103.001 & 8.51 \\
7713.830 & 8.04 & 8306.160 & 8.38 & 9122.115 & 8.99 \\
7723.681 & 9.43 & 8311.021 & 8.51 & 9128.115 & 7.99 \\
7724.642 & 9.14 & 8324.217 & 8.78 & 9147.110 & 8.56 \\
7728.690 & 8.42 & 8331.152 & 8.26 & 9150.087 & 7.80 \\
7735.672 & 8.36 & 8336.140 & 8.76 & 9160.098 & 8.75 \\
7740.667 & 9.09 & 8337.056 & 9.21 & 9183.034 & 8.92 \\
7741.774 & 9.08 & 8354.965 & 8.33 &          &      \\
7742.732 & 9.40 & 8359.182 & 8.80 &          &      \\
\hline
\end{tabular}
\end{center}
\end{table}

During this reduction procedure, possible inconsistencies were found among
Ichinohe's (1909) step reduction for JDs 2417949.206, 2417962.941,
2418390.133 and 2418413.047.  We consistently used our new values in the
following analysis.

After heliocentric corrections, the data were analyzed using the Phase
Dispersion Minimization (PDM) method (Stellingwerf 1978) to obtain the
pulsation period.  The resultant theta diagram is shown in Figure 1.
The best determined period of 22.18 d, which confirms the analysis
by Ichinohe (1909), well agrees with the 22.187 d period given before 1917
(Huth 1966).

\begin{figure}
  \begin{center}
  \includegraphics[angle=0,height=8cm]{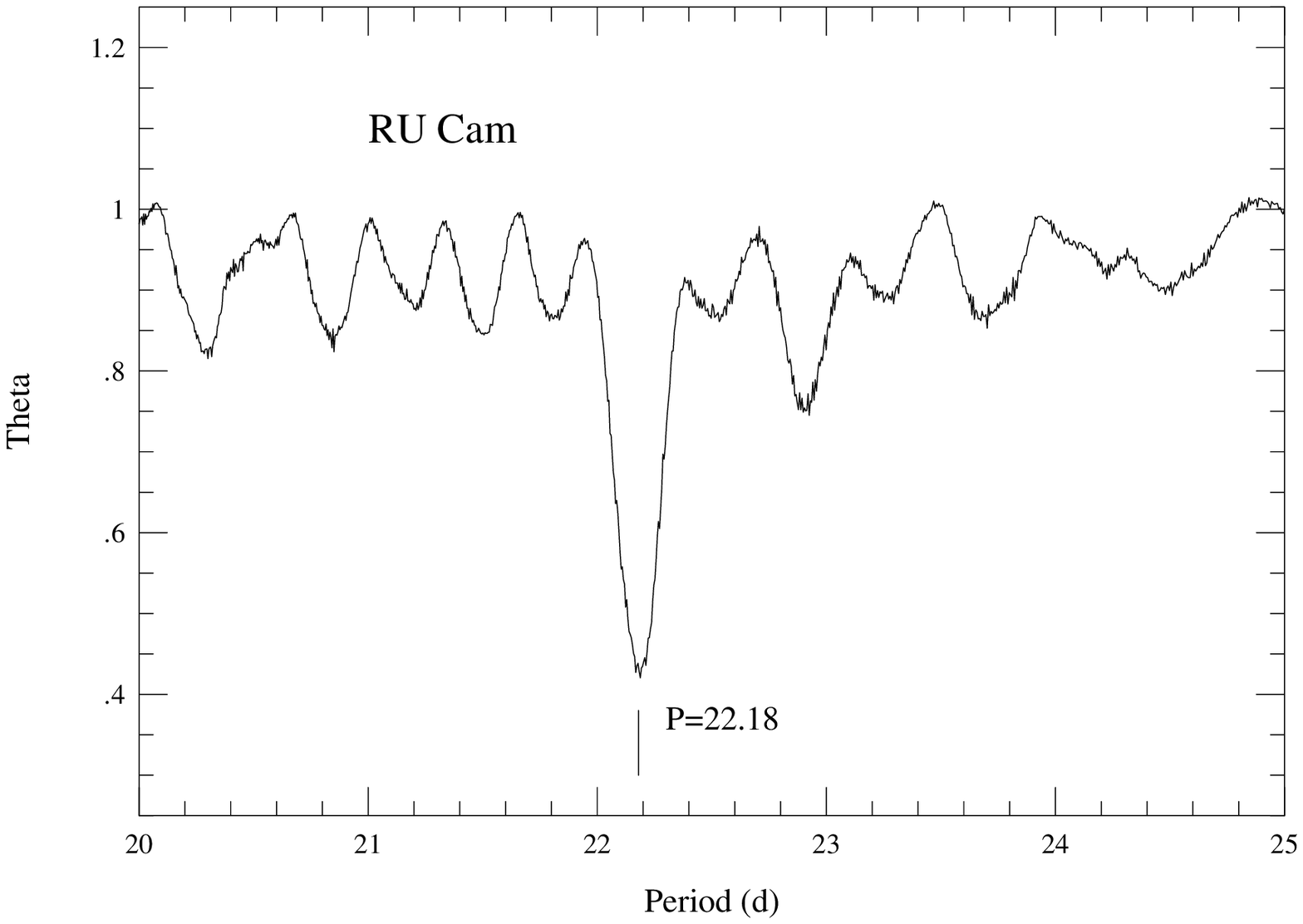}
  \caption{Period analysis of RU Cam} \label{fig-1}
  \end{center}
\end{figure}

The averaged light curve constructed from these observations is shown
in Figure 2.  The epoch of the light maximum was rather arbitrarily taken
as HJD 2419150.085.  Each point represents an average of 0.1-phase bin,
typically constituted of $\sim$ 10 observations, and vertical bar
1-sigma error.  The range of light variation was 8.26 ($\pm$ 0.06)
-- 9.33 ($\pm$ 0.08), giving an amplitude of 1.07 ($\pm$ 0.10), which
also confirmed the earlier reports.  Our new analysis, however, rather
disproves the existence of the secondary maximum which Ichinohe (1909)
described, giving a more consistent, typical W Vir-type, light curve
with those of other observers at that epoch.

\begin{figure}
  \begin{center}
  \includegraphics[angle=0,height=8cm]{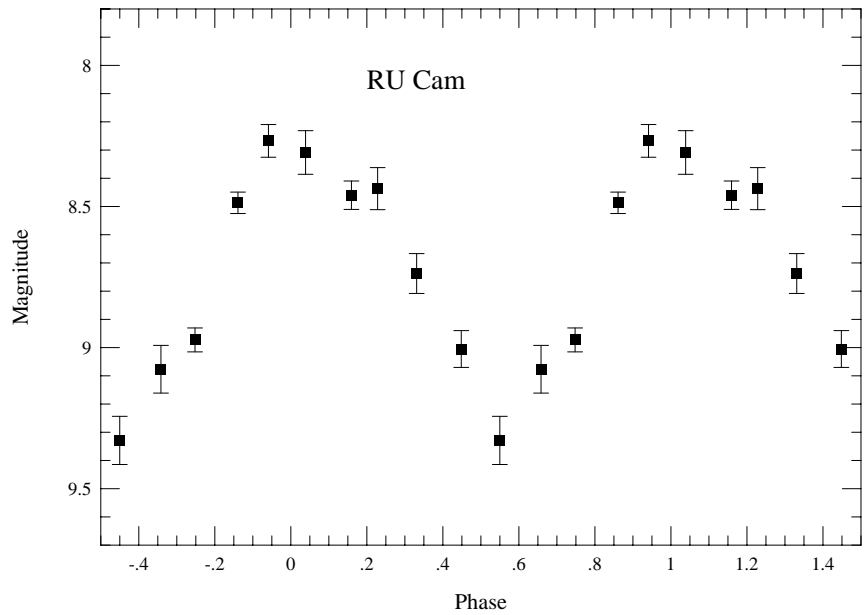}
  \caption{Averaged light curve of RU Cam} \label{fig-2}
  \end{center}
\end{figure}

\section*{Acknowledgments}
This work has been based on the VSOLJ Database.  The authors
are grateful to the VSOLJ staffs, and to M. Sato, who called our
attention to the early observations of RU Cam.

{}

\end{document}